\documentclass[reprint,footinbib,amsmath,amssymb,%
  aps,pre,floatfix,nolongbibliography]{revtex4-2}
\usepackage{graphicx}
\usepackage[colorlinks=true,allcolors = blue]{hyperref}

\newcommand{\gammadot}{{\dot\gamma}}
\newcommand{\Eq}[1]{Eq.~\eqref{#1}}

\newcommand{\Fig}[1]{Fig.~\ref{#1}}
\newcommand{\partFig}[2]{Fig.~\hyperref[#1]{\ref*{#1}#2}}

\newcommand{\naive}{na\"\i ve}
\newcommand{\role}{r\^ole}

\newcommand{\latin}[1]{{\itshape #1}}
\newcommand{\cf}{\latin{cf.}}

\newcommand{\eg}{\latin{e.\,g.}}
\newcommand{\etal}{\latin{et al.}}

\newcommand{\ie}{\latin{i.$\,$e.}}

\newcommand{\bonafides}{\latin{bona fides}}

\begin{document}

\title{Sliding friction in the hydrodynamic lubrication regime for
  a power-law fluid}

\author{Patrick B. Warren}

\email{patrick.warren@stfc.ac.uk}

\affiliation{Unilever R\&D Port Sunlight, Quarry Road East, Bebington,
  Wirral, CH63 3JW, UK.}

\altaffiliation{Current address: The Hartree Centre, STFC Daresbury Laboratory, Warrington, WA4 4AD, United Kingdom}

\date{February 4, 2015; July 9, 2023; published as J.\ Phys.:\ Cond.\ Matter {\bf29}, 064005 (2016).}

\begin{abstract}
A scaling analysis is undertaken for the load balance in sliding
friction in the hydrodynamic lubrication regime, with a particular
emphasis on power-law shear-thinning typical of a structured liquid.  It
is argued that the shear-thinning regime is mechanically unstable if
the power-law index $n<1/2$, where $n$ is the exponent that relates
the shear stress to the shear rate.  Consequently the Stribeck
(friction) curve should be discontinuous, with possible hysteresis.
Further analysis suggests that normal stress and flow transience
(stress overshoot) do not destroy this basic picture, although they
may provide stabilising mechanisms at higher shear rates.  Extensional
viscosity is also expected to be insignificant unless the Trouton
ratio is large.  A possible application to shear thickening in
non-Brownian particulate suspensions is indicated.
\end{abstract}


\maketitle


\section{Introduction} 
Lubrication is usually considered to be a mechanical engineering
problem \cite{Wil94, Ham94, Per00, Gre00, BVM02, ESB05}, demanding
sophisticated multi-scale numerical approaches for its
solution \cite{YTP06, CSR+07, JKB07, AK11, PS11}.  However lubrication
is also relevant for sensory physics, for example in understanding the
origins of `mouth-feel' and the sensory properties of
foods \cite{SMC+11, CS12}, and in `psycho-rheology' and the perception
of skin care products \cite{GMH+13}.  The lubricants in these sensory
physics applications are often structured liquids or soft solids, with
significant non-Newtonian rheologies.  This raises the question how
far traditional mechanical engineering lubrication theory can be used,
and indeed whether new phenomena may be encountered.  With this in
mind, I present a scaling analysis of sliding friction in the
hydrodynamic lubrication regime which extends to encompass the
power-law shear thinning behaviour commonly encountered in structured
liquids and complex fluids. The aim is to provide insights and
guidance for these unconventional application areas, and motivate
further work.  Appendix~\ref{app:squeeze} shows how the same approach can be
used to recover a number of known results in squeeze-flow
lubrication \cite{LB74, Rod96, LXH+01}.

\begin{figure}
\begin{center}
\includegraphics[width=0.95\columnwidth]{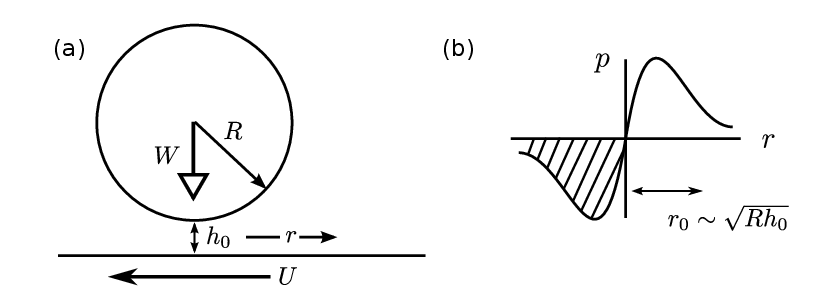}
\end{center}
\caption{Sliding friction between a sphere and flat.  (a) It is supposed that
  the flat surface is sliding at a velocity $U$ underneath the sphere.
  (b) The skew-symmetric Reynolds lubrication pressure
  distribution.\label{fig:geom} }
\end{figure}

\begin{figure}
\begin{center}
\includegraphics[width=0.95\columnwidth]{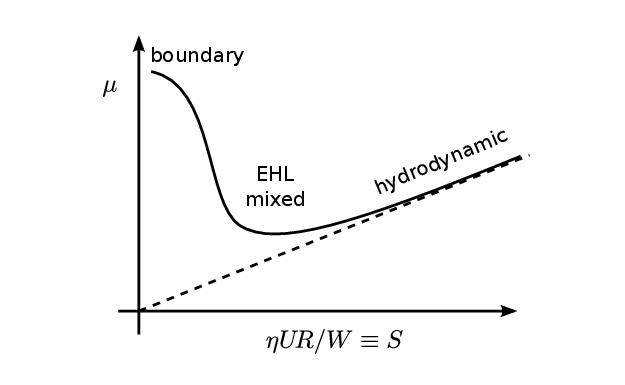}
\end{center}
\caption{Stribeck curve (friction coefficient as a function of
  Sommerfeld number).  EHL is elasto-hydrodynamic lubrication.  In the
  hydrodynamic regime (to which the present scaling analysis is
  applicable), the friction coefficient is proportional to the
  Sommerfeld number (dashed line).\label{fig:stribeck}}
\end{figure}

To begin with, let me outline the fundamental mechanism of pure
hydrodynamic lubrication in sliding friction \cite{Bat67, Fab95,
GHP+01}.  In a fully lubricated conjunction (\Fig{fig:geom}), mass
conservation within the converging and diverging wedges induces a
Poiseuille-like contribution to the entrainment flow, superimposed on
a Couette-type shearing motion.  There is a corresponding emergent
pressure distribution (\partFig{fig:geom}{b}), which supports the load.
The ratio of the load to the lateral sliding force defines the sliding
friction coefficient $\mu$.  As we shortly shall see, Reynolds
lubrication theory \cite{Wil94} predicts $\mu$ is proportional to
the \emph{Sommerfeld number} $S\equiv\eta U R/W$, where $\eta$ is the
lubricant viscosity, $U$ is the sliding velocity, $R$ is a length
scale characterising the curvature of the surfaces, and $W$ is the
normal load.  The overall $\mu(S)$ behaviour is often summarised in
the semi-empirical \emph{Stribeck curve} (\Fig{fig:stribeck}).  Note
that pure hydrodynamic lubrication breaks down as $S\to0$ since the
surfaces come into close contact, and one segues into a regime where
elasto-hydrodynamic lubrication (EHL) is relevant, and ultimately into
a regime of boundary lubrication (BL).  This is reflected in the
behaviour observed empirically in the Stribeck curve.  The present
scaling analysis does not incorporate EHL or BL, and should be
interpreted within this wider context.

Before embarking on the detailed development, one further general
remark should be made.  In a geometry which posesses reflection
symmetry, Reynolds lubrication theory predicts the lubrication
pressure should be skew-symmetric about the minimum gap
(\partFig{fig:geom}{b}).  Therefore, the integrated lubrication
pressure should vanish.  In actuality it has long been recognised that
some additional physics intervenes to knock down this result.  For
instance in the trailing edge where the lubrication pressure is
sub-ambient (shaded area in \partFig{fig:geom}{b}), the free surface
may separate \cite{Kap55}, or the fluid may cavitate \cite{AdPM05}.
These considerations make the exact solution to the problem dependent
on the nature of the additional physics.  However a simple and
commonly used prescription is to discard the contribution from the
sub-ambient pressure region---this is the so-called half-Sommerfeld
boundary condition \cite{Ham94}.  I shall tacitly use this assumption
below.

Keeping all this in mind, in the next section I shall develop scaling
arguments which recover the standard results for sliding friction in
the hydrodynamic lubrication regime.  In the following section I shall
extend these to power-law fluids.

\section{Scaling analysis: Newtonian case}
The gap between the lubricated surfaces, as shown in \partFig{fig:geom}{a}
for example, can be represented in the conjunction region by a
parabolic profile
\begin{equation}
h=h_0+\frac{r^2}{2R}\,.\label{eq:parab}
\end{equation}
In this $h_0$ is the minimum gap, $r$ is the radial distance from the
minimum gap, and $R$ is a measure of the radius of curvature of the
surfaces (\eg\ the sphere radius in \partFig{fig:geom}{a}).  The radius
within which $h$ remains of order $h_0$ defines a length scale
$r_0\sim ({Rh_0})^{1/2}$.  This length scale sets the size of the
conjunction region in terms of the Reynolds lubrication pressure
distribution (\partFig{fig:geom}{b}).  With this insight one can easily
calculate the mechanical properties of the conjunction such as the
supported load and the tangential friction force. Thus the
identification of $r_0$ is the key to the development of the scaling
argument.  This applies not only for sliding friction but also to
squeeze-flow lubrication as discussed in Appendix~\ref{app:squeeze}.

As already alluded to, and described in more detail in
Appendix~\ref{app:reynolds}, mass conservation induces a
Poiseuille-type flow of magnitude $U$ superimposed on the Couette-type
shearing motion.  The matching Reynolds lubrication pressure
distribution is such that $U\sim ({h_0^2}/{\eta})\nabla p$ where
$\nabla p$ represents the magnitude of the pressure gradient.  Note
that the lubrication pressure develops in the radial direction, thus
$r_0$ should be used as an estimate of the distance over which the
pressure varies (\partFig{fig:geom}{b}).  Denoting by $p$ the magnitude of
Reynolds lubrication pressure, the pressure gradient is therefore
estimated by $\nabla p\sim p/r_0$.  Putting these elements together, a
scaling estimate of the magnitude of the lubrication pressure is given
by
\begin{equation}
p\sim \frac{\eta U r_0}{h_0^2}\,.
\label{eq:p}
\end{equation}
As already outlined, the area of the conjunction in terms of the
Reynolds lubrication pressure distribution is $r_0^2\sim Rh_0$.  Then
the \emph{integrated} pressure (\eg\ the unshaded area in the
\partFig{fig:geom}{b}) corresponds to a normal force
\begin{equation}
p r_0^2\sim \frac{\eta U r_0^3}{h_0^2}\sim 
{\eta U R^{3/2}}{h_0^{-1/2}}\,.\label{eq:pr02}
\end{equation}
In this I have used $r_0\sim({R h_0})^{1/2}$ for the second step.
Note that this force diverges as $h_0^{-1/2}$ as the gap shrinks
($h_0\to0$).  The stable gap (see below) will be such that the
integrated lubrication pressure supports the load.  This happens when
$pr_0^2\sim W$, or alternatively, using \Eq{eq:pr02}, when
\begin{equation}
\frac{h_0}{R} \sim \Bigr(\frac{\eta U\!R}{W}\Bigl)^2\quad
\bigl({}\sim S^2\>\bigr)\,.\label{eq:gap}
\end{equation}
More careful treatments restore a prefactor to this, for example
Kapitza \cite{Kap55} derived $h_0/R=(72\pi^2\!/25)\,S^2\approx
28.4\,S^2$, and Hamrock \cite{Ham94} has $h_0/R\approx 34.8\,S^2$.

\begin{figure}
\begin{center}
\includegraphics[width=0.95\columnwidth]{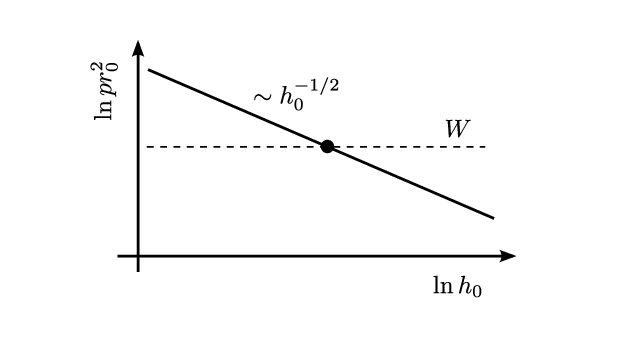}
\end{center}
\caption{Load curve in the Newtonian case.  The solid line is the
  normal force generated by the lubrication pressure.  The filled
  circle is a stable fixed point, under a given load.\label{fig:load}}
\end{figure}

A point not often stressed, but which will be critical in the sequel,
is that the lubricated conjunction is indeed \emph{mechanically
stable} at the load balance condition.  To see this suppose that the
minimum gap $h_0$ has not yet taken its steady state value.  The load
$W$ bears down on the conjunction, and is opposed by the lubrication
pressure which exerts a force of order $pr_0^2$ given
by \Eq{eq:pr02}.  However as already mentioned $pr_0^2\sim
h_0^{-1/2}$.  Therefore if the gap is too large, the lubrication
pressure does not support the load and the gap closes.  Conversely, if
the gap is too small, the lubrication pressure overcompensates for the
load and the gap opens up.  This stabilising mechanism is shown
schematically in \Fig{fig:load}, where the solid line is the normal force arising from the lubrication pressure.  In the language of dynamical systems
theory, the filled circle in this diagram is a \emph{stable fixed
  point}.

I turn now to the Stribeck curve.  Since the shear rate $\gammadot\sim
U/h_0$ the tangential wall stress in the conjunction zone is of order
$\eta\gammadot\sim\eta U/h_0$ (see Appendix~\ref{app:reynolds} for a
more careful justification of this).  Multiplying by the area of the
conjunction ($\sim r_0^2$) gives the tangential force $T \sim \eta
Ur_0^2/h_0$.  Hence the friction coefficient
\begin{equation}
\mu\equiv\frac{T}{W}\sim\frac{\eta Ur_0^2}{Wh_0}\sim \frac{\eta
  U\!R}{W}\quad\bigl({}\sim S\>\bigr)
\end{equation}
(using the geometrical condition $r_0^2\sim Rh_0$).  This now explains
why $\mu\propto S$ in the hydrodynamic part of the Stribeck curve.
Note that the gap shrinks as $h_0\sim S^2$ from
\Eq{eq:gap}.  As $S$ decreases (\ie\ low velocity, high load)
eventually the gap becomes very small and the friction coefficient
deviates from the pure hydrodynamic law at the point where one enters
the EHL/BL regime.\footnote{Repeating these arguments for a
two-dimensional geometry (\eg\ a cylinder and flat) finds $\mu\sim
S^{1/2}$ and $h_0/R\sim S$, where the Sommerfeld number $S=\eta
U/W_L$, and $W_L$ is the load per unit length.}

\section{Power-law fluids}
I now propose the obvious extensions of the above scaling arguments to
incorporate non-linear rheological effects of interest found in
structured liquids and complex fluids.  For the time being I shall
leave aside the question of normal stress, flow transience, and
extensional flow, and focus first on the typical power-law shear
thinning found in these systems.

\subsection{Shear thinning} 
I shall represent the non-linear viscosity in this case by a simple
model, \cf\ \cite{PS11},
\begin{equation}
\eta=\frac{\eta_0}{1+(\gammadot\tau)^\alpha}\,.\label{eq:eta}
\end{equation}
This model describes a power-law fluid with a low shear (first)
Newtonian plateau.  In \Eq{eq:eta} $\eta_0$ is the low shear
viscosity (\ie\ in the Newtonian plateau), $\gammadot$ is the shear
rate, $\tau$ is a characteristic relaxation time, and $\alpha$ is an
exponent characterising the rate of decrease of the viscosity in the
shear-thinning regime.  In the shear-thinning regime, the shear stress
$\eta\gammadot\sim \gammadot^{1-\alpha}$, thus we identify the
power-law index $n=1-\alpha$.  The analysis below will be undertaken
largely in terms of $\alpha$, but the results will also be presented
in terms of statements about the power-law index.  For concreteness, a
typical value of the exponent for polymer melt rheology is
$\alpha\approx0.8$--1 \cite{DE86, GLM+03, LG03} (power-law index
$0\alt n\alt0.2$).  One should note that $\alpha=1$ (which would
correspond to a plateau in the shear stress) is nowadays usually taken
to be a signal of shear banding \cite{DFM+16}.

The dimensionless quantity $\gammadot\tau$ in \Eq{eq:eta} is
the \emph{Weissenberg number} and is a measure of the extent to which
the non-linear regime has been penetrated.  \Eq{eq:eta} omits the
second Newtonian plateau which is expected to obtain at very high
shear rates; this will be added informally in a moment.

\begin{figure*}
\begin{center}
\includegraphics[width=0.8\textwidth]{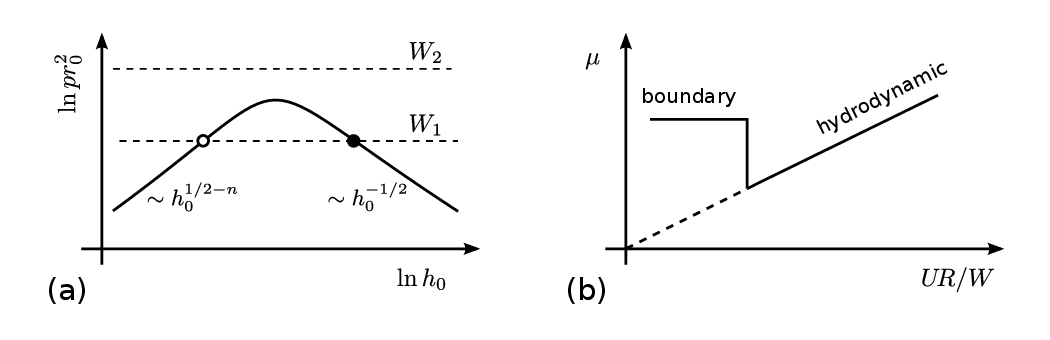}
\end{center}
\caption{(a) Load curves in the non-Newtonian strong shear thinning
  case (power-law index $n<1/2$).  The solid line is the normal
  force generated by the lubrication pressure.  A filled (open) circle
  is a stable (unstable) fixed point, under a given load.  The fixed
  points disappear in a saddle-node bifurcation as the load is
  increased.  (b) The corresponding Stribeck curve is predicted to
  jump discontinuously at the critical load, perhaps into a boundary
  lubrication regime.\label{fig:nonl}}
\end{figure*}

Inserting \Eq{eq:eta} into \Eq{eq:pr02} gives a revised estimate
for the normal force produced by the lubrication pressure,
\begin{equation}
pr_0^2\sim \frac{\eta_0}{1+(\gammadot\tau)^\alpha}\times
{U R^{3/2}}{h_0^{-1/2}}\label{eq:pr02x}
\end{equation}
where the shear rate is estimated by $\gammadot\sim U/h_0$.  The
assumptions being made here will be reviewed later.  Since
$\gammadot\sim 1/h_0$, there are two regimes.  For large gap, the
Weissenberg number is small ($\gammadot\tau\alt1$) and one
recovers the Newtonian behaviour seen already, with $\eta=\eta_0$
being the viscosity in the first Newtonian plateau.

For small gap, one enters the high
Weissenberg number regime ($\gammadot\tau\agt1$) in which case
\Eq{eq:pr02x} becomes
\begin{equation}
pr_0^2\sim
{\eta_0}{(\gammadot\tau)^{-\alpha}}\,
{U R^{3/2}}{h_0^{-1/2}}\sim h_0^{\alpha-1/2} \quad\bigl({}\sim h_0^{1/2-n}\>\bigr)\,.
\label{eq:pr02y}
\end{equation}
Only the $h_0$-dependence has been retained in the final step.  If
$\alpha<1/2$ (weak shear thinning, power-law index $n>1/2$), then
$pr_0^2$ in \Eq{eq:pr02y} diverges as $h_0\to0$ and one would again
expect the gap to be mechanically stable, albeit with a modified
dependence on the Sommerfeld number.  On the other hand, if
$\alpha>1/2$ (strong shear thinning, power-law index $n<1/2$),
$pr_0^2$ in \Eq{eq:pr02y} is an \emph{vanishing} function of $h_0$
as $h_0\to0$ and therefore the load balance in this regime will
be \emph{unstable}.

I shall take this latter case (strong shear thinning, power-law
index $n<1/2$) to be the one of most interest.  The two regimes in
the load balance are illustrated in
\partFig{fig:nonl}{a}.  The left hand branch is the high Weissenberg number
regime where shear thinning takes place.  In this branch, the open
circle corresponds to an unstable fixed point at load $W_1$.  To the
left of the open circle, the lubrication pressure is insufficient to
support the load and the gap closes until interrupted by some
additional physics.  To the right of the open circle, the lubrication
pressure forces the surfaces to move apart until one reaches the
stable fixed point (filled circle) in the low Weissenberg number
regime.  Now consider increasing the load to $W_2$.  In this
situation, the lubrication pressure is \emph{never} sufficient to keep
the surfaces apart and the gap closes until one enters the mixed or
boundary layer lubrication regime.  Somewhere in between $W_1$ and
$W_2$ is a critical load at which the stable and unstable fixed points
merge in a saddle-node bifurcation.

\begin{figure*}
\begin{center}
\includegraphics[width=0.8\textwidth]{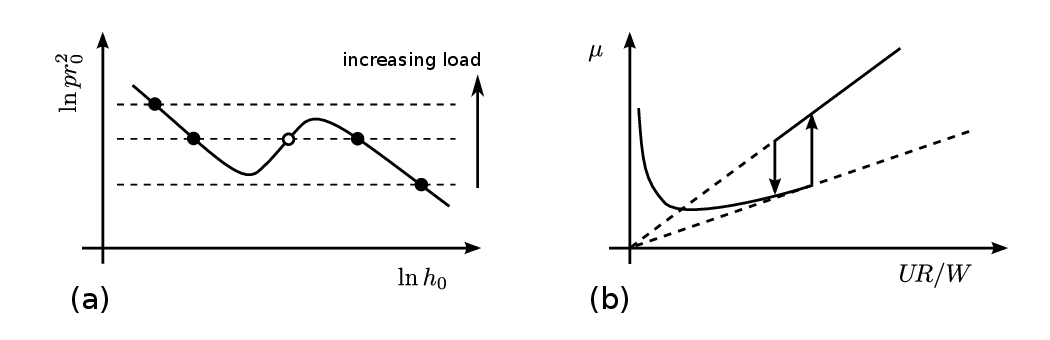}
\end{center}
\caption{(a) Lubrication pressure in the case of shear thinning with a
  second Newtonian plateau.  (b) Consequential hysteresis in the
  Stribeck curve.\label{fig:hyst}}
\end{figure*}

The implication for the Stribeck curve is as follows.  In the low
Weissenberg number regime, the analysis goes through as for the
Newtonian case, with $\eta_0$ featuring as the viscosity.  The
Weissenberg number $\gammadot\tau\sim S^{-2}$, and increases as the
Sommerfeld number shrinks.  Mechanical stability is lost at the point
where one enters the shear thinning regime (\ie\ at the saddle-node
bifurcation). At this point the surfaces should jump into (near)
contact.  Correspondingly the Stribeck curve should show a
discontinuous jump (\partFig{fig:nonl}{b}).

If one envisages that a second Newtonian plateau appears at very high
shear rates, then a restabilisation mechanism emerges naturally.  In
this situation, the expected behaviour of the lubrication pressure is
shown in \partFig{fig:hyst}{a}.  Consider increasing the load, starting
from a small value.  At first one follows the right hand branch of
the pressure curve, until stability is lost at the upper saddle-node
bifurcation and the system jumps onto the left hand branch.
Subsequently decreasing the load reverses the process, except that the
jump back occurs at the lower saddle-node bifurcation.  The
corresponding Stribeck curve displays \emph{hysteresis} and can be
constructed by noting that the stable fixed points lie on branches of
the lubrication pressure curve where the flow is Newtonian. This is
illustrated in \partFig{fig:hyst}{b}.

\subsection{Normal stress} 
Let me now consider some of the other non-linear rheological effects
that might be relevant for non-Newtonian lubricants.  A long standing
question which can be addressed in the context of the scaling analysis
is the \role\ of normal stress.  Again, a simple model capturing
this is
\begin{equation}
N=\frac{\psi_0\gammadot^2}{1+(\gammadot\tau)^\beta}\label{eq:norm}
\end{equation}
where $\psi_0\sim\eta_0\tau$ is a normal stress coefficient and
$\beta$ is another exponent.  For concreteness, a typical value of
this exponent is $\beta\approx 1.3$--1.5 for polymer melts
\cite{DE86, GLM+03, LG03}.

The contribution that normal stress makes to the load balance can be
found by multiplying by the area of the conjunction, $r_0^2\sim Rh_0$.
Alternatively we can just compare the normal stress with the Reynolds
lubrication pressure.  In the low Weissenberg number regime
($\gammadot\tau\alt1$) the ratio of these two is
\begin{equation}
\frac{N}{p}\sim \frac{\psi_0\gammadot^2}{\eta_0 Ur_0/h_0^2}
\sim \gammadot\tau\times \frac{h_0}{r_0}\,.
\end{equation}
where the pressure has been taken from \Eq{eq:p}, and $U\sim\gammadot
h_0$ and $\psi_0\sim\eta_0\tau$ have been substituted in the final
step.  The last factor $h_0/r_0\sim({h_0/R})^{1/2}$ is expected to be
a small number (\eg\ $\alt0.1$, say), so the ratio $N/p$ is expected
to remain small in the $\gammadot\tau\alt1$ regime.  In particular normal
stress should not affect the situation in the stable lubrication flow
up until the loss of stability at the saddle-node bifurcation in
\Fig{fig:nonl}.  Therefore the overall picture shown in
\Fig{fig:nonl} remains unchanged.

In the high Weissenberg number regime ($\gammadot\tau\agt1$) one
has $Nr_0^2\sim h_0^{\beta-1}$.  As long as $\beta>1$ this is an
increasing function of $h_0$, and the gap remains mechanically
unstable in this regime.  Only if $\beta<1$ will normal stress rescue
the conjunction from complete mechanical collapse by providing enough
normal force to support the load in a stable sliding configuration.
However since I have argued normal stress does not affect the loss of
mechanical stability at the saddle-node bifurcation, one would not
expect this stabilising mechanism to become relevant until well into
the non-linear regime.  Therefore, one would still expect hysteresis
in the Stribeck curve.

\subsection{Flow transience}
Another factor that can be considered is that the flow is transient on
a time scale of order the transit time $t_s\sim r_0/U$. The
dimensionless ratio $\tau/t_s\sim U\tau/r_0$ is known as the
\emph{Deborah number}, and its magnitude gives an indication of the
importance of transient flow effects, such as stress overshoot.
Substituting $U\sim\gammadot h_0$ shows that the Deborah number is of the
order $\gammadot\tau\times h_0/r_0$. This is the same as the ratio $N/p$
for the normal stress above, and the same argument goes through so
that the overall picture remains unchanged.  Like normal stress, it is
possible that transient flow effects may grow to become significant in
the non-linear regime since the Deborah number scales as $h_0^{-1/2}$.
Hence this may provide another mechanism to re-stabilise the gap at a
smaller distance.

\subsection{Extensional viscosity}
In Appendix~\ref{app:reynolds}, the ratio of the extension to the
shear components in the lubrication flow is estimated to be of the
order $(h_0/R)^{1/2}$.  This is the same small number encountered
above, and here should be compared to the \emph{Trouton ratio} (\ie\
between extensional and shear viscosity).  For example, for polymer
melts the Trouton ratio is of the order 3 in the low Weissenberg
number regime (where the gap is stable), and does not increase much in
the non-linear regime
\cite{DE86}.  On these grounds one would not expect extensional
viscosity to affect the friction curve, but the situation may have to
be re-evaluated if the Trouton ratio is large.

\section{Discussion}
The main result is that for sliding friction in the hydrodynamic
regime with a strongly shear thinning lubricant (power-law index
$n<1/2$) the conjunction is predicted to become mechanically unstable
at the point where the shear thinning regime is entered.  This is
because in the shear-thinning regime, with $n<1/2$, the integrated
Reyolds lubrication pressure \emph{diminishes} as the gap shrinks, and
is therefore unable to support the load.  The gap should therefore
collapse until some other physics intervenes.  This would lead to a
discontinuous jump in the Stribeck (friction) curve as the load is
increased (\partFig{fig:nonl}{b}).  Non-linear effects such as normal
stress and flow transience are estimated to be subdominant at the
point of entry into the non-linear regime, so it is unlikely they
would destroy the basic picture although they may provide stabilising
mechanisms at high shear rates.  Extensional viscosity is also
estimated to be insignificant unless the Trouton ratio is large (say,
$\agt10$).  Depending on the nature of the restabilising mechanism at
small gaps, the Stribeck curve may show hysteresis (\partFig{fig:hyst}{b}).

These results were established on the back of scaling arguments which
involve perhaps \naive\ and rather drastic assumptions.  For example,
the lubrication flow is assumed to remain essentially the same as in
the Newtonian case, but with a substituted shear-rate dependent
viscosity $\eta(\gammadot)$.  If this breaks down (for example due to
a non-linear superimposition of the Poiseuille and Couette flows) the
analysis would be invalid.  Furthermore, very simple models were taken
for the shear viscosity and normal stress as a function of shear rate,
and additivity of the normal stress and lubrication pressure was
assumed.  There are certainly available much more sophisticated
(tensor) constitutive models, such as the Rolie-Poly equation for
polymer melts \cite{LG03}, or the phenomenological White-Metzner
model \cite{WM63}.  An important avenue for future work is to
investigate more thoroughly the properties of Reynolds lubrication
flow using such constitutive models.  An intriguing possibility is
that the non-linear rheology could break the reflection symmetry in
the lubrication pressure profile (\partFig{fig:geom}{b}), thus relieving
the theory of the necessity to appeal to some unstated physics to
generate a resultant normal force.

The possible hysteresis in the Stribeck curve makes an interesting
connection with recent theories of the microscopic origins of shear
thickening in non-Brownian suspensions \cite{FMR+13, SMM+13, WC14}. In
these theories the shear thickening transition is associated with a
`breakthrough' to frictional contacts, and indeed the simulations in
Fernandez \etal\ \cite{FMR+13} are based on a discontinuous friction
curve of exactly the form shown in \partFig{fig:nonl}{b}.  Of course, the
non-Brownian particles that show the phenomenon are more typically
suspended in a Newtonian solvent, and not a structured liquid, and
therefore one can question the relevance of the present analysis.
Nevertheless, experimental systems are often prepared with polymer
additives (as in \cite{FMR+13}), and as such they may perhaps
exhibit non-linear frictional behaviour of the kind described here if
significant polymer adsorption occurs.

I thank M.\ J.\ Adams and S.\ A.\ Johnson for useful discussions, and
critical reading of early versions of the manuscript.

Note added (July 2023): in relation to this problem what is commonly called `normal' stress is actually manifest as a tensile `hoop' stress (as in the rod-climbing effect) and so does not \emph{directly} contribute to supporting a load in a sliding geometry.  This re-inforces the conclusion that normal stress cannot stop a conjunction from becoming unstable in sliding, but makes the mechanism by which it could rescue the conjunction from complete collapse at high Weissenberg number rather indirect: a full study remains an open problem to my knowledge. I thank Alexander Morisov for drawing my attention to this point.

\appendix

\section{Aspects of Reynolds lubrication flow}\label{app:reynolds}
For a conjunction between non-conformal surfaces
(\cf\ \partFig{fig:geom}{a}), mass conservation implies that entrainment
involves a superimposition of Couette and Poiseuille flows.  The full
solution requires numerics \cite{Ham94} but basic insights can be
gained by considering the quasi-one-dimensional case \cite{GHP+01}.
In that case the flow is
\begin{equation}
v_x =
-\frac{U}{h}\,(h-y)-\frac{1}{2\eta}\frac{dp}{dx}\,y(h-y)\,.\label{eq:a1}
\end{equation}
Here $x$ (\cf\ $r$) measures the distance along the gap, $y$ measures
the distance from the lower surface (moving at velocity $-U$), $h$ is
the gap (weakly varying with $x$), $p$ is the Reynolds lubrication
pressure, and $\eta$ the viscosity.  This velocity field corresponds
to a net material flux
\begin{equation}
Q= \int_0^h\!\! v_x\,dy = -\frac{Uh}{2}-\frac{h^3}{12\eta}\frac{dp}{dx}\,.
\end{equation}
This has to be constant, even though $h$ varies, and is of order
$Uh_0$.  The implication is that $(h_0^3/12\eta)\,dp/dx\sim U h_0$,
which gives an order of magnitude estimate of the pressure gradient.
The associated Poiseuille flow velocity (along the centerline for
instance) is then $(h_0^2/\eta)\, dp/dx\sim U$, as utilised in the
main discussion.

The shear rate at the lower surface is 
\begin{equation}
\frac{\partial v_x}{\partial y}\Big|_{y=0} = 
\frac{U}{h}-\frac{h}{2\eta}\frac{dp}{dx}\,.\label{eq:a3}
\end{equation}
Since $dp/dx\sim \eta U/h_0^2$, the two terms are comparable, and the
wall stress can be estimated by $\eta U/h_0$.  This give rise to the
friction force in hydrodynamic lubrication.

The flow field in \Eq{eq:a1} is predominantly in the longitudinal
direction.  Differentiating with respect to this direction yields an
estimate for the extensional component.  This yields a number of
terms, all of which are of the order $U/h\times dh/dx$.  Since
$h=h_0+x^2/2R$ (\cf\ \Eq{eq:parab}), one has $dh/dx = x/R$, and
setting $x\sim x_0\sim(h_0R)^{1/2}$ and $h\sim h_0$ gives $U/x_0$
for the magnitude of the extensional strain rate.  This agrees with
the simple picture that entrainment involves of the order 100\%
extensional strain, in a distance of the order $x_0$, on a time scale
of the order the transit time $x_0/U$.  Hence the extensional strain
rate $U/x_0\sim (U/h_0)\times (h_0/x_0)\sim \gammadot\times  (h_0/R)^{1/2}$.
This is the origin of the estimate used in the main text.

\begin{figure}
\begin{center}
\includegraphics[width=0.95\columnwidth]{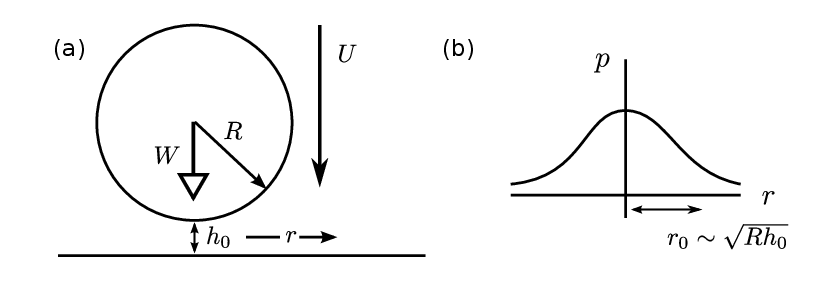}
\end{center}
\caption{Squeeze-flow lubrication between a sphere and flat.
(a) The sphere approaches the flat surface at a velocity $U$.  (b) The
  radially-symmetric Reynolds lubrication pressure
  distribution.\label{fig:squeeze} }
\end{figure}

\section{Squeeze-flow lubrication}\label{app:squeeze}
Scaling arguments along the lines of the main text can also be
developed for squeeze-flow lubrication.  In this way one recovers
number of known results \cite{LB74, Rod96, LXH+01, ESB05}, which may
help prove the \bonafides\ of the approach.  The main difference
between squeeze-flow and sliding friction lies in the estimate of the
radial flow rate $u_r$.  Here, mass conservation in the conjunction
zone (\partFig{fig:squeeze}{a}) means that $U r_0^2\sim u_r r_0 h_0$, and
therefore $u_r\sim r_0 U/h_0$ (\cf\ $u_r\sim U$ for the sliding
friction case).  Thus in squeeze-flow lubrication the radial flow rate
is amplified by a factor $r_0/h_0$ compared to sliding case.  From
here, the development proceeds exactly as in the main text.  The
radial flow corresponds to a pressure gradient such that
$(h_0^2/\eta)\nabla p\sim u_r\sim r_0U/h_0$.  The pressure gradient is
estimated in terms of the magnitude of the Reynolds lubrication
pressure by $\nabla p\sim p/r_0$.  Finally, the normal force is
estimated by $pr_0^2$.  When all this is put together,
\begin{equation}
p r_0^2\sim \frac{\eta r_0^4 U}{h_0^3}\,.
\end{equation}
Although `\emph{sans}' prefactor, this is actually a very old result
due to Stefan \cite{ESB05}.  At this point we have not yet used the
geometric relation $r_0\sim\sqrt{Rh_0}$.  If this is inserted, we
obtain
\begin{equation}
p r_0^2\sim \frac{\eta R^2 U}{h_0}\,.
\end{equation}
This is the classic Reynolds squeeze-flow lubrication force; a more
careful treatment yields the exact result that the normal force
$F=3\pi\eta R^2U/2 h_0$ \cite{Ham94}.  Note that the Reynolds
lubrication pressure is radially symmetric in this case
(\partFig{fig:squeeze}{b}), and there is no need to appeal to hidden
physics as in the half-Sommerfeld boundary condition.

The shear rate in the gap is estimated as $\gammadot\sim u_r/h_0\sim
r_0U/h_0^2\sim R^{1/2}U/h_0^{3/2}$.  For power-law shear-thinning with
$\eta\sim\gammadot^{-\alpha}$ we find (omitting the intermediate
steps) the normal force
\begin{equation}
pr_0^2\sim h_0^{3\alpha/2-1}\quad\bigl({}\sim h_0^{(1-3n)/2}\>\bigr)\,.
\end{equation}
If $\alpha>2/3$ (power-law index $n<1/3$), this no longer diverges as
$h_0\to0$.  This signals that the force is controlled by the
outer-region flow (on the length scale $\sim R$).  This rather
striking result was first reported by Rodin \cite{Rod96}, and later
confirmed by Lian \etal\ \cite{LXH+01}.  In a sense, this is the
analogue of the main text observation that for sufficiently strong
shear thinning, the conjunction fails to support a load.  The
difference in the cross-over exponent value can be traced to the
geometric amplification of the radial flow rate in the present
situation.

The effect of normal stress can be assessed in the same way.  Only the
main results are summarised.  First, in the low Weissenberg number
regime one concludes that normal stress will be small compared to the
Reynolds lubrication pressure.  Second, in the high Weissenberg number
regime one finds that $N/p\sim h_0^{3(\beta-\alpha)/2-1}$.  This
diverges as $h_0\to0$ if $\beta<\alpha+2/3$.  This is a distinct
possibility for the exponent values typical of a structured liquid and
would mean that normal stress \emph{is} relevant for squeeze-flow
lubrication in the high Weissenberg number regime.  A more detailed
investigation of this possibility is left for future work.

%

\end{document}